\documentclass[lettersize,journal]{IEEEtran}
\usepackage{amsmath,amsfonts}
\usepackage{algorithmic}
\usepackage{algorithm}
\usepackage{array}
\usepackage[caption=false,font=normalsize,labelfont=sf,textfont=sf]{subfig}
\usepackage{textcomp}
\usepackage{stfloats}
\usepackage{url}
\usepackage{verbatim}
\usepackage{graphicx}
\usepackage{cite}
\begin{document}

\title{Harnessing Large Language Models for Intelligent Resource Allocation in the Internet of Everything}

\author{Haijun Zhang,~\IEEEmembership{Fellow,~IEEE}, Zhuojun Duan, Zijun Wu, Xu Ma, Yuzheng Ren
\thanks{Haijun Zhang, Zhuojun Duan, Zijun Wu, Xu Ma, and Yuzheng Ren are with Hebei Key Laboratory of Space-Air-Ground Intelligent Communication and Beijing Engineering and Technology Research Center for Convergence Networks and Ubiquitous Services, University of Science and Technology Beijing, Beijing, China, 10083 (email: zhanghaijun@ustb.edu.cn, m202410612@xs.ustb.edu.cn, wuzijun@xs.ustb.edu.cn, maxu@xs.ustb.edu.cn, renyuzheng@ustb.edu.cn)}
}

\maketitle

\begin{abstract}
The rapid development of the Internet of Everything (IoE) is accelerating the adoption of intelligent applications. However, the massive number of connected devices generates diverse and heterogeneous tasks, which pose increasing challenges for dynamic resource scheduling in IoE environments. Using their superior semantic understanding and reasoning capabilities, Large Artificial Intelligence Models (LAIMs) demonstrate significant potential to handle complex scheduling scenarios and improve resource utilization efficiency. This paper investigates a task-oriented LAIM-driven resource scheduling mechanism, which constructs a multidimensional scheduling decision model by integrating task semantics, network states, and constraint conditions. Furthermore, a task-oriented prompt generation method is designed to establish a deep association between task requirements and network state. In the proposed resource allocation scheme, an external evaluation and feedback module is incorporated to conduct real-time feasibility verification and performance evaluation of scheduling strategies, thus enhancing the robustness and adaptability of scheduling. Simulation results demonstrate that the proposed Large Language Model (LLM)-driven network architecture and resource allocation scheme achieve significant improvements in convergence speed, processing latency, and energy consumption, effectively enhancing IoE task responsiveness and resource utilization.
\end{abstract}


\section{Introduction}
\IEEEPARstart{T}{he} breakthroughs of Large Artificial Intelligence Models (LAIMs) in semantic understanding, contextual reasoning, and knowledge generalization have significantly expanded the capabilities of parameter-driven intelligent systems \cite{shen2024large}. Using large-scale data sets for training and fine-tuning in multi-task scenarios, these models demonstrate strong generality and adaptability, and have been widely applied in natural language processing, multimodal perception, and automated decision-making \cite{zhang2025resource}. As the scale of models continues to grow and the pre-training paradigm keeps evolving, LAIMs are advancing from perceptual intelligence to cognitive intelligence, becoming a key driving force for intelligent transformation across diverse domains. Supported by next-generation networking and computing infrastructures, LAIMs are gradually penetrating complex application scenarios such as the industrial Internet, smart cities, and autonomous driving, providing solutions for highly dynamic data management and diversified decision-making \cite{tang2025towards}.

As a key component of the next-generation information network, the IoE is facing unprecedented challenges in resource management. IoE systems deploy a wide range of terminal devices and edge nodes to support diverse tasks such as industrial sensing, environmental monitoring, autonomous driving, and immersive interaction. These tasks exhibit distinct resource requirements, including high computational density, significant bandwidth fluctuations, strong dependence on caching, and sensitivity to energy consumption \cite{9574695,huang2023resources}. Against this background, designing refined resource scheduling mechanisms based on task-specific real-time requirements and service objectives has become an important research direction that drives the evolution of resource management \cite{zhang2023toward}. Traditional intelligent algorithms are capable of learning optimal strategies in unsupervised environments. However, when addressing task-oriented dynamic resource allocation, they often fail to capture the complex coupling between task semantics and network states in a timely manner. This limitation leads to significant challenges for traditional intelligent algorithms in terms of convergence speed and generalization capability, especially in highly dynamic and constraint-intensive IoE environments \cite{du2020green}. Therefore, it is urgent to introduce intelligent mechanisms with strong generalization and knowledge-based reasoning capabilities. Such mechanisms can enable efficient perception of dynamic task requirements, accurate modeling of constraints, and coordinated optimization of multi-objective scheduling \cite{jamil2022resource}.

Large Language Models (LLMs) demonstrate strong capabilities in natural language understanding, complex knowledge reasoning, and context-aware modeling \cite{liu2024large}. These capabilities make LLMs an ideal foundation for building intelligent scheduling architectures in IoE environments. In particular, in environments with highly dynamic resource demands, complex task semantics, and diverse optimization objectives, LLMs can demonstrate the ability to sense and guide global policy objectives \cite{lee2024llm}. Recent studies have leveraged LLMs to coordinate specialized Artificial Intelligence (AI) modules and Internet of Things (IoT) devices for executing complex tasks \cite{cui2024llmind}. This approach addresses the limitation of traditional IoT systems, which struggle to complete cross-device tasks through natural language instructions efficiently. Reference \cite{tan2025tool} proposed a tool-aided evolutionary LLM. In this framework, the LLM selects devices based on natural language prompts, while convex optimization tools jointly solve CPU frequency, transmission power, and bandwidth allocation in real time. The system evolves strategies offline using a GPT-2-based virtual environment combined with population-based relative policy optimization, thereby enabling efficient energy management for wireless federated learning in dynamic environments. Although LLMs have demonstrated superior performance in resource management compared with traditional Reinforcement Learning (RL) methods \cite{he2024large}, many researchers are also exploring hybrid approaches that combine RL with LLMs to achieve better results \cite{yao2025enhancing}. For example, the authors in \cite{tang2025llm} designed a resource scheduling framework based on the Gemma 2B model. They constructed an expert knowledge base fine-tuned on optimization objectives, which generated dynamic action candidate sets to guide the decision-making of RL agents.

Although the above methods have demonstrated the potential of LLMs in resource optimization tasks, existing research still faces certain limitations. Most current approaches only use LLMs for static prompting or reward shaping, without establishing a systematic closed loop from semantic prompting to policy feedback. This limitation restricts the full exploitation of the adaptive optimization capabilities of LLMs. In addition, the ability of LLMs to handle complex system constraints, multi-task scheduling conflicts, and resource heterogeneity modeling has not yet been comprehensively evaluated or systematically analyzed. To address these issues, this paper proposes a novel architecture for IoE resource management. The architecture integrates three components: a semantic-driven prompt generation engine based on LLMs, an RL agent deployed at the edge, and an external evaluator that provides real-world constraint verification and reward estimation. These three components collaborate to dynamically perceive environmental states and jointly optimize computational resources, caching resources, and wireless bandwidth resources.

The main contributions of this article can be summarized as follows:
\begin{itemize}
    \item[$\bullet$] A resource optimization architecture that integrates LLMs, RL, and an evaluation-feedback mechanism is proposed. This architecture combines the semantic reasoning capability of LLMs with the reward-driven policy optimization of RL, thereby achieving strong generalization and task adaptability.
    \item[$\bullet$] A task-driven semantic prompting and reward-shaping mechanism is designed. By incorporating a semantic prompt-driven policy search process, this mechanism improves the convergence speed and generalization capability of RL-based methods.
    \item[$\bullet$] An external evaluation module is introduced to conduct fine-grained verification and feedback on policy actions, which enhances system training efficiency and stability.
\end{itemize}

The remainder of this article is organized as follows. Section II introduces the LLM-driven IoE network architecture. Section III describes the proposed LLM-guided Prompt Reinforcement Optimizer (LLM-PRO) in detail. Section IV presents the simulation results and provides a discussion. Section V outlines several open issues and challenges. Finally, Section VI concludes the paper.

\begin{figure*}[!t] 
	\centering 
	\includegraphics[width=0.8\textwidth]{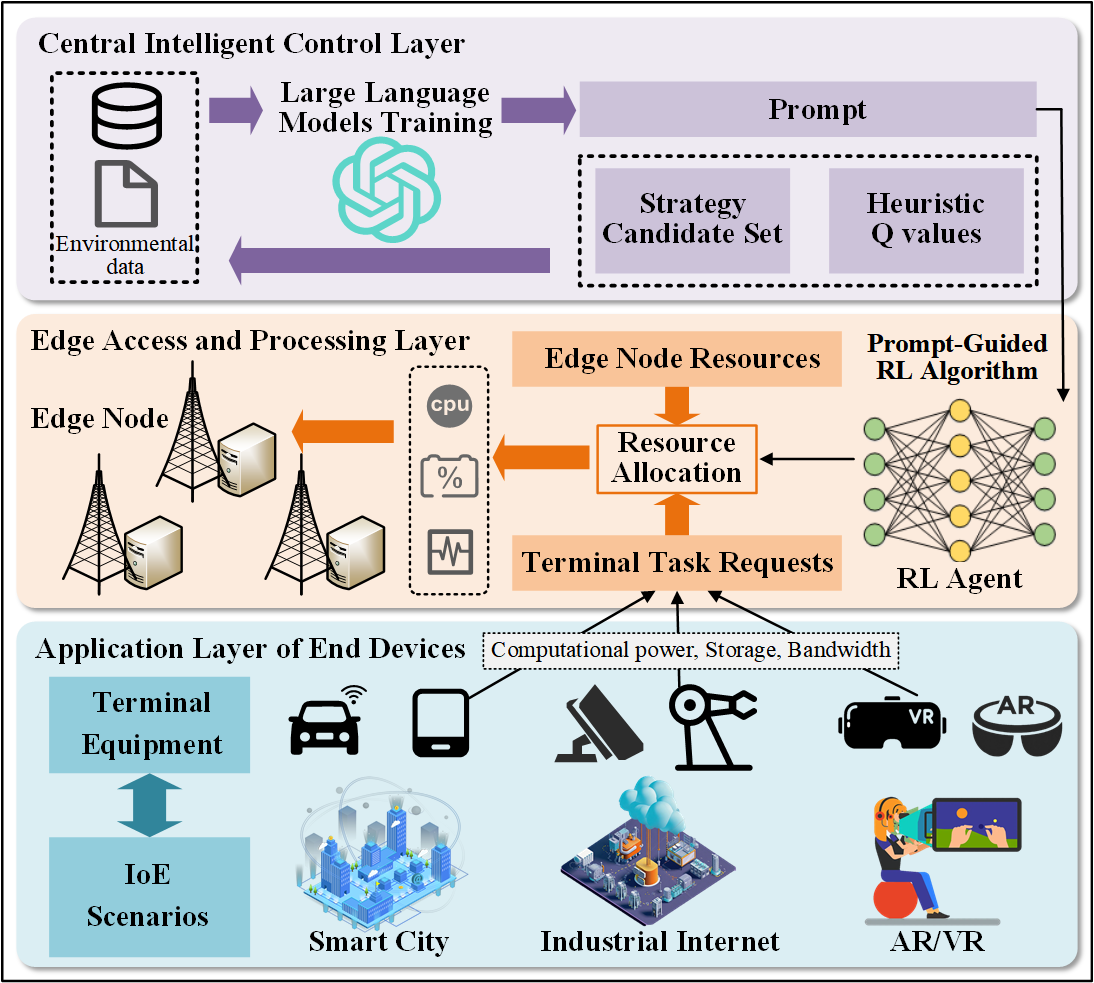}
	\caption{Large Language Model-Driven Network Architecture for IoE Resource Scheduling.} 
	\label{fig_1} 
\end{figure*}

\section{LLM guided IoE resource allocation}
The IoE is gradually moving toward deep integration of intelligent sensing, ubiquitous computing, and dynamic communications. As a result, resource scheduling faces new challenges with higher levels of dynamism and complexity. LLMs provide new opportunities for addressing these challenges, as they possess strong capabilities in understanding complex scenarios, rapidly learning and adapting to dynamic rules, and performing global optimization under multidimensional constraints. Fig. 1 illustrates the proposed LLM-driven IoE resource scheduling network architecture. The architecture adopts a multilayer collaborative design to support resource scheduling functions. The system is divided into three functional layers, which correspond to three key stages: task requests, resource scheduling execution, and global strategy guidance. The network relies on language-prompt-driven mechanisms combined with self-learning strategies to schedule computing and communication resources in edge computing scenarios dynamically.
\subsubsection*{\bf Application Layer of End Devices}
A wide range of IoE devices are extensively deployed, including mobile terminals, industrial sensors, vehicular platforms, and unmanned systems. Each terminal dynamically initiates task requests based on its service requirements and periodically reports its environmental status. The task types exhibit significant heterogeneity, typically including latency-sensitive AR/VR applications, throughput-oriented data stream tasks, and periodic data collection tasks with specific energy constraints. The resource demands of these devices vary dynamically, directly affecting network scheduling strategies and quality of service control.
\subsubsection*{\bf Edge Access and Processing Layer}
Several edge nodes with real-time decision-making capabilities are deployed, typically implemented as base station edge servers on the operator side. These nodes are equipped with heterogeneous multi-core processing units, high-speed local caching modules, and wireless resource control modules. They can perform localized resource evaluation and scheduling responses for the tasks they serve. Each node integrates a lightweight RL agent, which constructs state vectors and leverages policy guidance from the central prompting module. By doing so, it continuously optimizes task scheduling strategies and enables efficient management and on-demand allocation of edge resources.
\subsubsection*{\bf Central Intelligent Control Layer}
With the computing support of cloud data centers, a pre-trained LLM is deployed as the semantic engine for system-level policy generation. Its role is to perform semantic modeling of global task descriptions and node states, and to generate policy prompts and heuristic reward functions aligned with optimization objectives. The LLM periodically receives state summaries and task information reported by edge nodes. Based on this input, it produces structured prompts that include target preferences, constraint conditions, and heuristic reward functions. These prompts are transmitted back to edge nodes to guide reward construction and policy fine-tuning in local RL processes. Unlike static policies, the LLM has the capability to fuse information across tasks and nodes. It can dynamically adapt to environmental changes and feedback during the policy convergence process. By continuously adjusting prompt content, the LLM enhances the system’s generalization and robustness in dynamic environments.

In an IoE network, the challenge of resource scheduling primarily arises from the large number of intelligent terminals. These terminals generate task requests to the network side based on their service scenarios, task characteristics, and real-time states. Each request consists of computational, storage, and bandwidth requirements. These requests reflect the resources that terminals rely on from edge nodes during local task execution. Accordingly, the edge access processing layer must efficiently allocate and dynamically regulate resources to handle each task request. The set of edge nodes is distributed across the entire edge access processing layer. Each node differs in terms of CPU performance, remaining memory, available storage space, and allocatable bandwidth. Moreover, different request-to-node mappings and task execution orders can have varying impacts on network performance indicators, such as task processing latency, node energy consumption, and load distribution at the edge layer. Therefore, leveraging LLMs as tools to achieve fine-grained mapping between large-scale task requests and edge processing nodes in IoE networks is the key focus of the resource management approach studied in this paper. 

\begin{figure*}[!t] 
	\centering 
	\includegraphics[width=1   \textwidth]{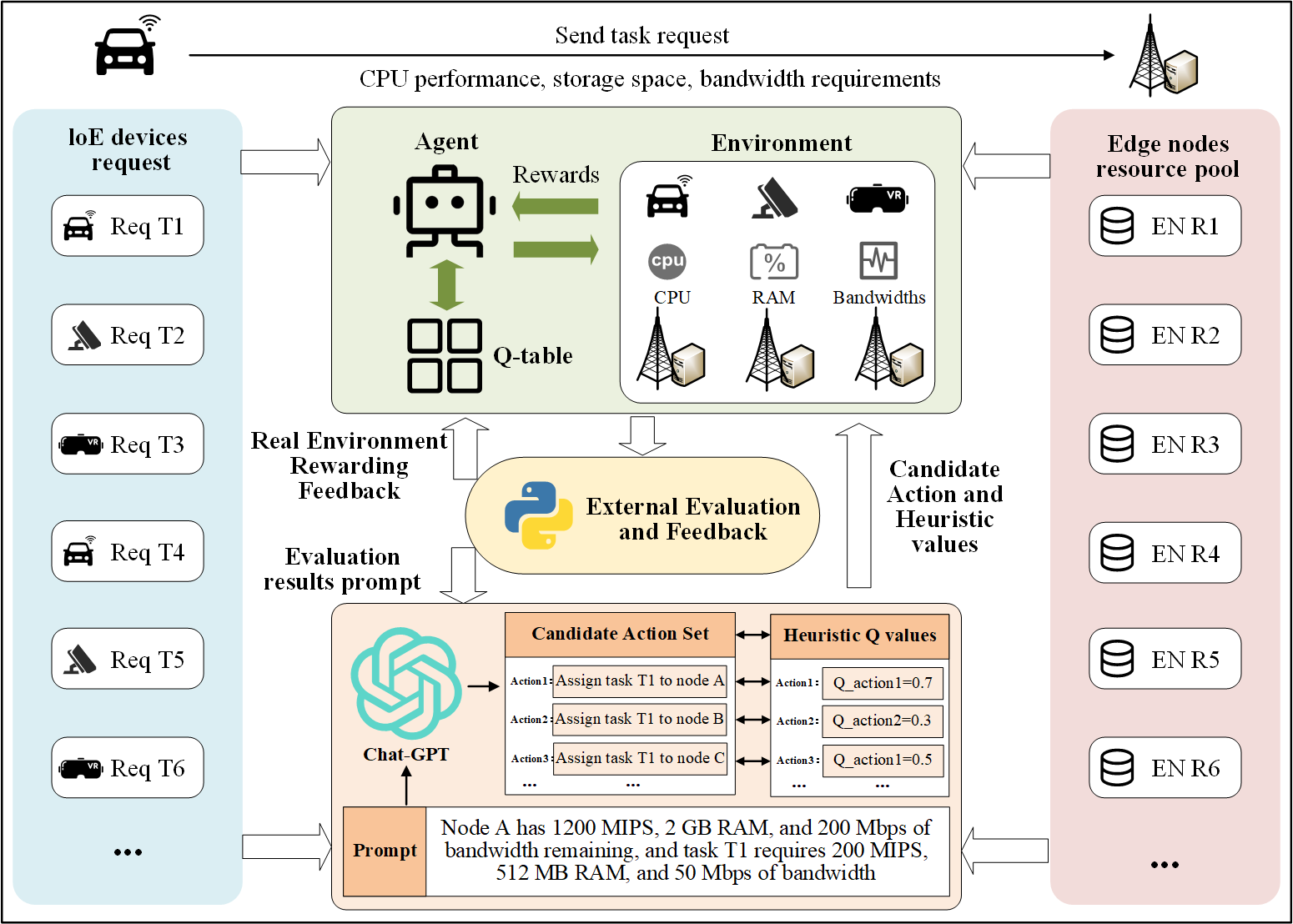}
	\caption{LLM-driven IoE resource scheduling network architecture.} 
	\label{fig_2} 
\end{figure*}
\section{LLM-guided Prompt Reinforcement Optimizer}
This paper proposes the LLM-PRO, whose basic architecture is illustrated in Fig. 2. LLM-PRO is designed to enable intelligent, scalable, and context-aware resource scheduling in large-scale IoE environments. The scheduling mechanism consists of three components: an LLM, an RL agent, and an external evaluation and feedback module. The framework operates on a multi-cycle collaborative iteration mechanism, which generates policy prompts, executes localized decisions, and continuously feeds back policy performance.
\subsection{LLM-Based Prompt and Evaluation-Driven Resource Optimization Architecture}
In the proposed LLM-PRO, the LLM mainly performs two core functions: semantic prompt generation and heuristic reward shaping. Specifically, the LLM periodically receives natural language prompts generated from environmental states and task requests. These prompts contain information such as the remaining MIPS, RAM, and bandwidth of each edge node, as well as the computational, memory, storage, and bandwidth requirements of each task request. The prompts are constructed using predefined Python string templates, which render numerical fields into natural language sentences. For example,'Node A has $1200$ MIPS, $2$ GB RAM, and $200$ Mbps bandwidth remaining. Task $T1$ requires $200$ MIPS, $512$ MB RAM, and $50$ Mbps bandwidth.' Based on the input prompts, the LLM outputs a heuristic Q-value vector for each candidate action set, thereby explicitly linking the semantic description to a specific task-to-node scheduling decision. These Q-values are not derived from actual execution results but are instead prior estimations informed by task types and environmental semantics. In this way, they provide directional guidance signals for the RL agent during cold-start or exploration phases. The heuristic Q-value vector is directly returned to the RL agent and combined with immediate reward shaping terms and the Temporal-Difference targets in Bellman updates. No additional inference from the LLM is required until the next LLM update cycle triggers a new prompt generation and vector output.

As the update cycles of the LLM continue, the external evaluation and feedback module reports the actual rewards of the actions executed by the RL agent as well as any constraint violations back to the LLM. This feedback is used to improve the LLM’s contextual understanding and prompt generation strategy. After each RL agent training round, a parsing script is automatically invoked to convert the evaluation results into a structured summary. The summary includes the influence of the current heuristic Q-value vector on agent behavior, the deviation from actual rewards, and whether constraint violations occurred. This information is appended to the following prompt as additional context, forming an external loop that enables adaptive learning of the LLM.

\subsection{Semantic Prompt-Guided Reinforcement Learning Algorithm}
To improve the efficiency of resource scheduling decisions and the context adaptation capability of the agent, a Q-learning (QL) architecture enhanced with an LLM is constructed within LLM-PRO, as illustrated in Fig. 3. By combining language-prompt-driven mechanisms with policy search, this architecture enables high-quality optimization of multidimensional resource scheduling strategies.

The QL agent constructs its input based on both the environmental state and the heuristic Q-value vector generated by the LLM for each candidate action. The Q-values generated by the LLM are not directly used. Instead, they are combined with the Q-values computed in real time by the Q-network through weighted fusion, producing a fused Q-value vector. This fused vector serves as the primary basis for the agent’s action selection. The agent adopts an $\epsilon$-greedy strategy to select an action from the fused Q-value vector. The chosen action is then executed, which triggers a state transition in the environment and yields an immediate reward. This reward is derived from performance indicators such as task processing latency and node energy consumption, which are calculated based on the actual network state. Since the LLM's Q-values influence action selection during the fusion process, the resulting reward also implicitly reflects the contribution of the LLM’s heuristic guidance to reward shaping. The agent still follows the core mechanism of the RL paradigm. Under the experience replay mechanism, the fused reward signal, together with the current state, action, and subsequent state, is written into the experience buffer. The agent samples historical interaction trajectories from the buffer for batch training and updates the policy network parameters through the Bellman equation, thereby continuously improving its resource allocation capability. In our framework, the LLM is updated on a slower planning cycle and only regenerates the heuristic Q-value vector when the system state changes significantly, while the RL agent makes multiple fast scheduling decisions between two LLM updates using the latest heuristic, whose weight decays exponentially over training so that early exploration is LLM-guided and later control gradually reduces to pure QL until convergence. Therefore, when the LLM is unavailable, the agent falls back to a purely local decision mode, where policy exploration is carried out by its own RL updates combined with heuristic rules.

\begin{figure}[!t] 
	\centering 
	\includegraphics[width=3.5in]{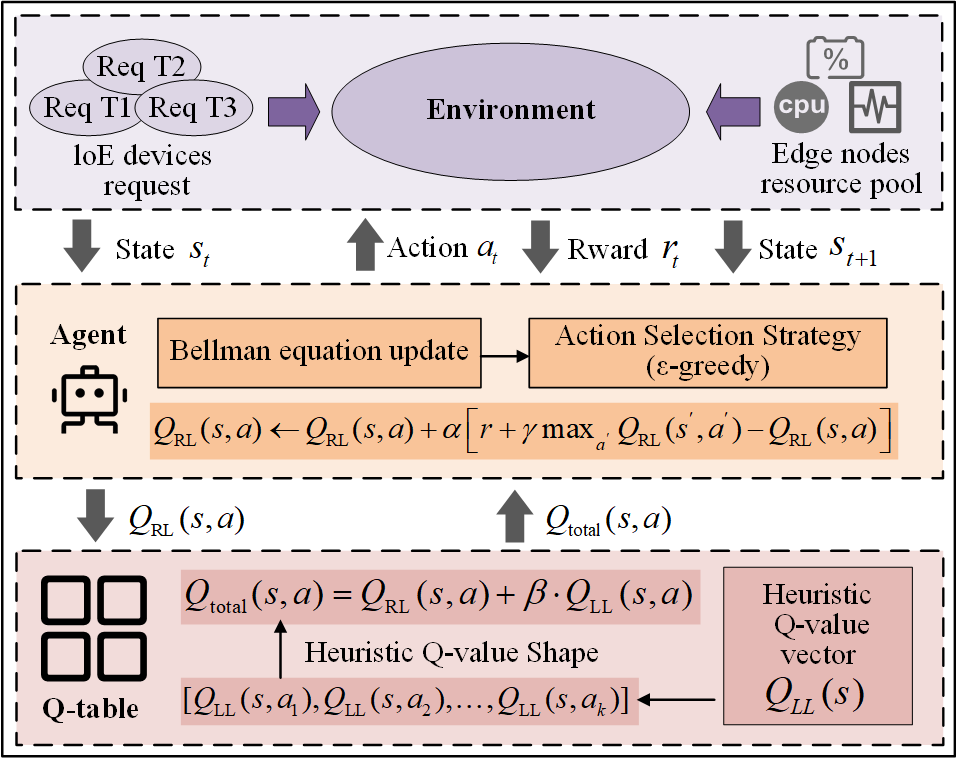}
	\caption{LLM-enhanced QL architecture.} 
	\label{fig_3} 
\end{figure}

\subsection{External Evaluation and Feedback}
In LLM-PRO, an external evaluation and feedback module is introduced as a key component for performance verification and closed-loop feedback. This module validates the feasibility of scheduling policies generated by the RL agent in real environments, evaluates rewards, and provides contextual enhancement support. It thus offers high-quality and trustworthy feedback for collaborative system optimization.

During each interaction round, the module receives the scheduling actions produced by the RL agent together with the current environmental state. Based on this input, it performs constraint verification and performance estimation. The evaluator first checks the consistency of resource allocation schemes in the scheduling actions. Using information such as network link status, cache occupancy, and node computing capability, it verifies whether the scheduling plan violates hard system constraints, including maximum concurrent task limits, spectrum conflict constraints, cache overflow, and energy budgets. Once a violation is detected, the system immediately returns violation notifications and assigns explicit negative feedback to guide policy convergence toward the feasible solution space. If the scheduling plan satisfies all constraints, the evaluator further models and analyzes its execution performance. By applying data-driven micro-simulation, it reconstructs several key performance indicators, including average latency, node energy consumption, and edge-layer load. Based on this analysis, it generates a high-precision reward value that reflects real-environment feedback.

Moreover, the results from the evaluator are also appended in textual form to the LLM’s context. This enables the LLM to reference historical strategy outcomes and violation records in the next round of semantic prompt generation. Through this mechanism, the LLM not only produces more targeted semantic rewards and guiding prompts, but also dynamically expands its knowledge boundary and continuously improves the quality of its prompts.

\subsection{Complexity and Overhead Discussion}
In the proposed architecture, the RL agents are executed locally at the edge nodes, while the centralized LLM is only queried periodically to update the heuristic Q-value vector. As a result, the per-decision complexity at each edge node is identical to that of standard QL, which means it requires $O(|A|)$ operations to evaluate the fused Q-values, where $|A|$ denotes the size of the candidate action set. The additional complexity introduced by the LLM is dominated by its inference on the aggregated state summaries, which scales with the number of edge nodes rather than the number of end devices. Therefore, the proposed LLM-PRO preserves low decision latency at the edge while leveraging cloud-side semantic guidance.

\section{Simulation Results}
This section aims to validate the performance of the proposed LLM-PRO in IoE environments through numerical results. Comparative experiments are conducted on the same simulation platform with the same parameter settings against traditional QL, the Particle Swarm Optimization (PSO) algorithm, and the Deep Q-Network (DQN) method, which are widely used baseline approaches for resource management in IoE and edge computing scenarios. An IoE scenario encompassing 50 end devices and 5 edge nodes is constructed to simulate a medium-scale IoE edge computing environment characterized by resource-constrained edge nodes. Each end device generates computation tasks according to an independent Poisson process and offloads them to one of the edge nodes according to the scheduling policy. The CPU capacity of each edge node is randomly drawn from 800 to 1600 MIPS, and the RAM is drawn from 2048 to 4096 MB to emulate heterogeneous but resource-constrained edge devices. For simplicity, the uplink transmission rate is set to a constant value of 100 Mbps, which corresponds to a typical IoE uplink configuration with 20 MHz bandwidth. The experimental platform integrates the ChatGPT-4o model as the LLM-guided policy generation and guidance module. As a representative large language model, ChatGPT-4o provides heuristic Q-values to the RL agent, thereby accelerating the exploration process of the algorithm. In the experiment, the large language model is invoked once every 50 training steps. Between invocations, the reinforcement learning agent reuses the most recent heuristic Q-value vector and updates its policy based on environmental rewards. To keep the cost bounded, the length of each prompt is restricted by encoding only aggregate statistics of the edge nodes and a small set of candidate actions, with the maximum number of generated tokens capped to 256 per call.

\begin{figure}[!t] 
	\centering 
	\includegraphics[width=3.5in]{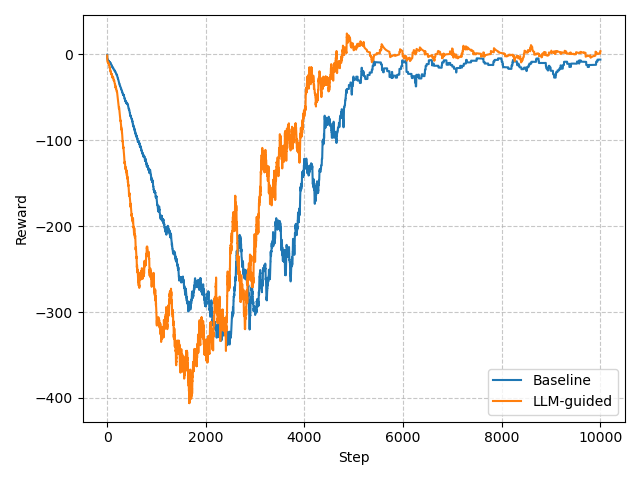}
	\caption{Convergence curves of rewards for LLM-PRO and traditional QL.} 
	\label{fig_4} 
\end{figure}
Fig. 4 presents the comparison of convergence curves for the average cumulative rewards during training between LLM-PRO and traditional QL. QL exhibits large fluctuations in the early iterations and requires a more extended exploration period compared with LLM-PRO. In contrast, LLM-PRO demonstrates a clear upward trend at an early stage and achieves faster convergence. This indicates that, with LLM guidance, the agent can establish a high-quality state–action mapping in the initial phase of policy search, enabling it to learn effective resource allocation strategies more quickly and significantly reducing the required training cycles. It is worth adding that, according to experimental observations, each LLM inference introduces approximately 4 KB traffic, and the end-to-end latency of a single ChatGPT-4o API call on our machine is about 0.8 second on average, which is negligible compared to the total training time. ChatGPT-4o is used only in the offline training phase to shape the initial exploration of the RL agent. After convergence, the learned policy is executed at the edge without any LLM queries. Therefore, the LLM inference latency does not affect the real-time performance of online IoE resource scheduling. Moreover, the average reward achieved by LLM-PRO in the stable phase is also notably higher than that of QL, which further confirms that the policy quality obtained by LLM-PRO is superior to traditional QL methods.
\begin{figure}[!t] 
	\centering 
	\includegraphics[width=3.5in]{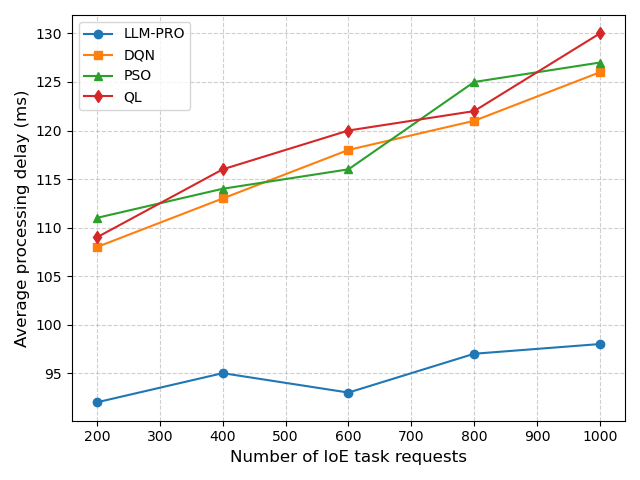}
	\caption{Relationship between the number of IoE task requests and average task processing latency.} 
	\label{fig_5} 
\end{figure}

\begin{figure}[!t] 
	\centering 
	\includegraphics[width=3.5in]{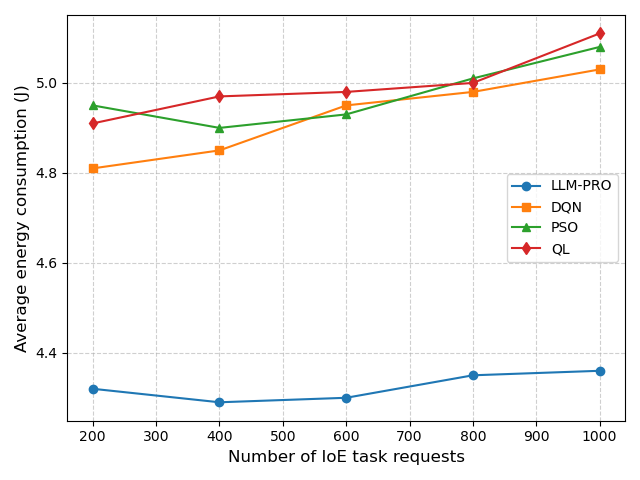}
	\caption{Relationship between the number of IoE task requests and average task processing energy consumption.} 
	\label{fig_6} 
\end{figure}

Fig. 5 compares the average task processing latency of the proposed method with that of QL, PSO, and the DQN. The experiment investigated the impact of varying numbers of IoE task requests on average task processing latency, whilst maintaining a fixed number of terminal devices and edge nodes. As shown in the figure, LLM-PRO consistently achieves the lowest average latency under different task request volumes. This improvement is attributed to the LLM’s semantic understanding of task requests, which guides the agent to learn optimal Q-values. In addition, the external evaluation and feedback module provides real network feedback that prevents irrelevant information from interfering with scheduling decisions, thereby enabling efficient matching between task resource demands and node resource allocation. As the number of task requests increases, the other three algorithms exhibit a clear growth in latency. In contrast the latency variation of LLM-PRO remains relatively stable across different workload conditions. This stability results from the LLM-guided RL agent’s ability to rapidly match tasks with the most suitable nodes based on task characteristics and network states, thus reducing queuing delays. These results fully verify the effectiveness and robustness of the proposed LLM-PRO in edge node resource allocation.

Fig. 6 compares the average task processing energy consumption of the proposed method with that of QL, PSO, and the DQN. The experiment investigates the impact of the number of IoE task requests on average task processing energy consumption. As shown in the figure, LLM-PRO consistently achieves lower average energy consumption than the other three methods. This advantage results from LLM-PRO’s ability to account for the matching degree between task characteristics and node states in scheduling decisions, thereby avoiding unnecessary task migrations and frequent use of high-power nodes. In contrast, the other three methods generate a large number of inefficient allocations during policy search, which leads to frequent high-power operation of edge nodes. Moreover, under high-load conditions, the energy consumption of these methods increases rapidly, whereas LLM-PRO maintains stable node energy consumption even in such conditions. The experimental results demonstrate that LLM-PRO not only reduces the total amount of energy consumption but also leverages the external evaluation and feedback mechanism to avoid frequent ineffective scheduling. Therefore, it achieves more reliable resource allocation in IoE multi-user scenarios.

\section{Open Issues and Challenges}
In the LLM-driven resource management framework for IoE environments, preliminary studies have demonstrated its great potential in task semantic understanding, optimization prompt generation, and scheduling strategy guidance. However, several key challenges remain to be addressed before its deployment in diverse and dynamic network environments.

{\bf{Challenges in Multi-source Information Fusion and Semantic Modeling:}}
Scheduling problems in IoE environments often involve complex information inputs that come from heterogeneous sources, have varying granularity, and change dynamically over time. This imposes higher demands on the semantic understanding and contextual reasoning capabilities of LLMs. Existing models are mostly pre-trained on general-purpose corpora and lack domain-specific knowledge of wireless system states and scheduling logic. As a result, the generated prompts may deviate from actual physical constraints and task requirements. Therefore, designing a dynamic semantic alignment mechanism that integrates network knowledge with environmental feedback to enhance the understanding of spatiotemporal dynamics and resource correlations is an important research direction for achieving accurate resource scheduling.

{\bf{Deep Integration of Semantic-driven and Predictive Scheduling:}}
Current IoE scheduling largely relies on observed system states, which can lead to delayed resource allocation in fast-changing network environments. By combining AI-based LLMs with environmental prediction models, it becomes possible to infer future network states and task demands in advance and perform proactive resource pre-allocation based on the predictions. This approach can reduce scheduling latency and significantly improve resource utilization. However, under multi-source heterogeneous data with uncertainty and noise, ensuring prediction accuracy and stability of prompt generation, as well as enabling rapid correction of scheduling strategies when predictions are inaccurate, remains a major technical bottleneck for this integration.

{\bf{Deployment Challenges of LLMs in Low-power Edge Environments:}}
Current architectures primarily deploy LLMs on the cloud side to fully exploit centralized computing and storage capabilities. However, with the growing demand for real-time responsiveness and distributed intelligence, deploying LLMs at edge nodes has become a critical approach to improve system responsiveness and reduce backhaul latency. Edge nodes, however, are generally constrained in terms of computing power, memory, and energy consumption, making it highly challenging to operate large-scale LLMs efficiently. Thus, how to ensure inference accuracy and contextual understanding while reducing overhead through model compression, pruning, parameter-efficient tuning, operator optimization, and heterogeneous computing collaboration is a core technical challenge for the sustainable deployment of LLMs on low-power platforms. Moreover, distilling LLMs into lightweight models and adapting them to IoE workloads via parameter-efficient fine-tuning is also an important means to realize lightweight LLM deployment at edge nodes.

{\bf{Integration with Advanced DRL and LLM-assisted Scheduling:}}
In the current design, we employ a QL-based agent enhanced by LLM guidance, rather than more sophisticated DRL algorithms. Although modern deep reinforcement learning methods, such as Deep Deterministic Policy Gradient (DDPG), Proximal Policy Optimization (PPO), and Rainbow DQN, as well as hybrid schemes that combine them with heuristic rules, offer significant advantages in high-dimensional continuous control scenarios, directly deploying such high-complexity algorithms on IoE edge nodes remains challenging. Running actor-critic DRL algorithms with deep neural networks, large replay buffers, and frequent parameter updates on every edge device would quickly exhaust the available computational budget. How to integrate advanced DRL algorithms such as DDPG, PPO, and Rainbow DQN into the LLM-PRO framework therefore remains a key problem that we plan to address in future work.

\section{Conclusion}
Resource scheduling driven by large AI models is an important research direction for future IoE systems. This paper proposes a task-oriented LLM-driven resource allocation scheme, which builds a multidimensional scheduling decision model by integrating task semantics, network states, and constraint conditions. To ensure efficient utilization of multi-source heterogeneous information, a prompt generation mechanism based on LLMs is designed to dynamically associate task requirements with network environments, while RL is employed to achieve adaptive optimization of resource allocation strategies. In addition, an external evaluation and feedback module is introduced to conduct real-time feasibility verification and performance assessment of scheduling policies. The evaluation results are further fed back into the prompt generation and policy optimization processes, ensuring sustained decision effectiveness under complex environments. Simulation results demonstrate that the proposed scheme effectively improves network resource utilization. The LLM-driven framework is expected to be a key enabling technology for achieving reliable resource scheduling in IoE scenarios such as smart manufacturing, intelligent transportation, and green energy management.

	\bibliographystyle{IEEEtran}
	\bibliography{reference.bib}

\end{document}